\documentclass[preprint]{aastex}
\def\hide#1{}

\begin{document}
\title{Correlation between Hard X-ray Peak Flux and Soft X-ray Peak Flux 
in the Outburst Rise of Low Mass X-ray Binaries}
\author{Wenfei Yu\altaffilmark{1}, Michiel van der Klis\altaffilmark{2}, 
and Rob Fender\altaffilmark{2} }
\altaffiltext{1}{Center for Theoretical Astrophysics and Department of 
Physics, 241 Loomis Lab, University of Illinois at Urbana-Champaign, 
Urbana, IL 61801. E-mail: wenfei@uiuc.edu}
\altaffiltext{2}{Astronomical Institute, ``Anton Pannekoek'', University 
of Amsterdam, Kruislaan 403, 1098 Sj Amsterdam, The Netherlands. E-mail: 
michiel,rpf@science.uva.nl}

\begin{abstract}
We have analyzed {\it Rossi} X-ray timing explorer (RXTE)
pointed observations of the outbursts of black hole and neutron 
star soft X-ray transients in which an initial low/hard state or 
`island' state, followed by a transition to a softer state, 
was observed. In three sources, the black hole transient 
XTE J1550-564, the neutron star transient Aquila X-1 
and a quasi-persistent neutron star low mass X-ray binary 
(LMXB) 4U 1705-44, two such outbursts were found. 
We find that the flux of the soft X-ray peak, 
which lags the hard X-ray peak by a few days to 
several weeks, scales with the flux of the hard X-ray peak. 
We conclude that we are able to predict the soft X-ray 
outburst peak flux based on the `preceding' hard X-ray 
peak flux, implying an early set up of the outbursts. We also 
find that the X-ray luminosity corresponding to the peak 
of the hard X-ray flux, which corresponds to the X-ray luminosity of 
the start of the hard-to-soft state transition, varies by a factor 
of about 2. If the accretion geometry early in the outburst rise 
is composed of two flows 
(e.g. a hot sub-Keplerian halo flow and a Keplerian disk flow, or an 
outflow and a Keplerian disk flow), 
the correlation indicates that the two flows are initially 
related, probably due to processes in the outer part of the 
accretion disk. We discuss constraints on a single flow model 
and a disk-jet model from these observations.
\end{abstract}


\section{Introduction}
When a black hole soft X-ray transient(SXT) evolves from quiescence 
to an outburst, it may go through a series of state
transitions, from a low/hard state in the rising phase of their
outbursts to an intermediate state, a high/soft state, or even a
very high state, which of these softer states a source reaches 
is probably related to the maximal mass accretion rate that
an outburst achieves. For neutron star low mass X-ray binary 
(LMXB) system, the spectral/timing state transition is shown in 
the color-color diagram (Hasinger \& van der Klis 1989) as movement 
from `island' state to `banana' state, with additional complicated 
movement toward/from lower X-ray luminosity 
(Muno, Remillard \& Chakrabarty 2002; Gierlinski \& Done 2002; 
Barret \& Olive 2002). In both neutron star and black
hole system, {\it an apparent hard flare will be observed if there is 
an initial low/hard state associated with a state transition 
in the outburst rise}. Such hard flares were observed in the 
outbursts of black holes XTE J1550-564 (Wilson \& Done 2001; Hannikainen
et al. 2001; Dubath et al. 2003), 4U 1630-47 (Hjellming
et al. 1999), XTE J1859+226 (Brocksopp et al. 2002; Hynes et al.
2002), and GRO 0422+32 (Ling et al. 2003), and the neutron stars 
Cen X-4 (Bouchacourt et al. 1984) and Aquila X-1 (Yu et al. 2003), 
suggesting a common origin of state transitions among black
hole and neutron star SXT outbursts. The hard
flare precedes the soft X-ray maximum by an interval of several days 
to more than ten days, corresponding to the time span 
of the hard-to-soft state transition and the rise of a soft state. 
The {\it most luminous} low/hard state occurs in the early outburst 
rise at the peak of the 
hard `preceding' flare for an individual outburst. 
This {\it most luminous} low/hard state is the best target to probe 
the spectral/timing properties of the low/hard state because of its 
high X-ray flux (Yu et al. 2003). 

The rise and the decay of an outburst has been 
explained in terms of the accretion
disk approaching and receding from the compact object. 
Such a picture is likely directly related to the
hysteresis effect of state transitions during an outburst rise or
decay (Miyamoto et al. 1994), where hard-to-soft and soft-to-hard
state transitions are different -- one corresponds to the disk approaching 
the center and the advection dominated accretion flow (ADAF) 
(Narayan \& Yi 1996) or the corona collapses, 
and the other corresponds to disk receding from the center perhaps 
because of disk evaporation 
(Meyer \& Meyer-Hofmeister 1994; Meyer et al. 
2000). 

The above accretion geometry composed of a 
truncated accretion disk and a hot spherical ADAF or corona at 
the center has been challenged by the long-term monitoring 
observations of some black hole X-ray binaries, which favor 
an accretion geometry containing two accretion flows 
(Smith, Heindl \& Swank 2002). In this paper, we show 
evidence supporting two accretion flows {\it related in \.{M}}. 
Our results  also indicate that the X-ray luminosity 
corresponding to the hard-to-soft state transition during 
the outburst rise varies by a factor of 2.   

\section{Observations and Results}
Using the same {\it Rossi} X-ray Timing Explorer (RXTE) 
light curve retrieval method we described in Yu et al. (2003), 
we analyzed the Proportional Counter Array (PCA) and the High Energy 
X-ray Timing Explorer (HEXTE) pointed observations of some 
black hole and 
neutron star soft X-ray transients (SXTs) in the public archive before 
May 2003. We follow Yu et al. (2003) to obtain daily average 
ASM (2-12 keV) light curves and HEXTE (15-250 keV) light curves, 
and found outbursts with a low/hard state preceding 
the soft X-ray maximum, in the 
following sources -- the black hole X-ray binaries  
XTE J1550-564 (JD 2451065 and 2451628), 
XTE J1859+226 (JD 2451460), 4U 1630-47 (JD 2450868), XTE J1650-500 
(JD 2452158) and the neutron star X-ray binary 
Aquila X-1 (JD 2451310 and 2451810), and in the flares of the 
neutron star LMXB 4U 1705-44 (JD 2451235 and JD 2451370). We focus on 
XTE J1550-564, Aquila X-1 and 4U 1705-44, the {\it only} sources 
in which we found more than one outburst/flare (i.e. two) 
showing that the hard X-ray peak 
precedes the soft X-ray peak (see below).   

\subsection{ASM and HEXTE Daily Average Light Curves}

\paragraph{XTE J1550-564} The 1998-1999 and the 2000 April outbursts 
are shown in Fig.~1 (left) and Fig.~1 (right), respectively. The 
hard X-ray observation of the 1998 outburst with BATSE has been 
reported in Hannikainen et al. (2001) and Wu et al. (2002), 
while BATSE detection of the 2000 outburst was reported in 
Masetti \& Soria (2000) and Jain \& Bailyn (2000). For the 
1998-1999 outburst, the peak HEXTE flux of the `preceding' hard flare 
is 468$\pm$1 c/s. The peak ASM flux we use in the following analysis 
is 140 c/s, estimated by excluding the very high state (VHS) interval 
marked by the dash-dotted lines for reasons explained below. 
We estimate the uncertainty of this flux to be $\pm$10 c/s based on the 
variation in ASM count rate during the primary outburst plateau. 
For the 2000 outburst, the peak HEXTE count rate 
recorded was 256$\pm$1 c/s (the true peak flux may have been 
270 c/s, see Fig.~1 right) and the peak ASM rate is 70$\pm$1 c/s. 
Here the count 
rate error are rounded up to the nearest integers unless 
otherwise indicated. We exclude the VHS because we want to consistently compare the 
black hole soft X-ray 
transient XTE J1550-564 with neutron star X-ray binaries, in which 
the characteristics of the VHS, i.e. a simultaneous 
high soft X-ray flux and high hard X-ray flux, have never been observed. 

\paragraph{Aquila X-1} We already reported that in the 1999 outburst 
and the 2000 outburst of Aquila X-1, the hard X-ray flare precedes 
the soft X-ray outburst (Yu et al. 2003). The peak HEXTE count rates recorded are 
56$\pm$1 c/s and 105$\pm$1 c/s, and the peak ASM count rates are 29$\pm$1 c/s 
and 50$\pm$2 c/s, respectively. The HEXTE pointing cover 
the hard peak of the 1999 outburst but 
may not cover that of the 2000 outburst which by 
extrapolation may have reached about 120 c/s, 15 c/s higher 
than the observed. 

\paragraph{4U 1705-44} 4U 1705-44 is not a transient source. However, its long-term 
variability and detailed spectral evolution suggest that it probably undergoes 
the same state transitions as SXTs. In
Fig.~2, we show ASM and HEXTE light curves regarding the two flares of 
the source. The observed HEXTE peak count rates are 34$\pm$1 c/s and 
48$\pm$1 c/s, and the peak ASm count rates are 21$\pm$1 c/s and 31$\pm$1 c/s, respectively. 
 
\subsection{Correlation between the HEXTE Peak Flux and the ASM Peak Flux}
In Fig.~3, we plot the {\it observed} HEXTE peak flux vs. the ASM peak 
flux for the outbursts or flares of XTE J1550-564, Aquila X-1, 4U 1705-44, and 
several other sources. The uncertainties 
of peak flux because of lack of coverage are very small. 
The plot shows a proportionality 
between the hard X-ray peak flux and the delayed soft X-ray peak 
flux for each source. The data corresponding 
to the three sources are also roughly consistent with a single 
linear proportionality, indicating similar {\it observed} energy spectra 
among these sources. Deviation from the proportionality may come 
from different Galactic or intrinsic absorption among sources.   

Because the energy spectra corresponding to the hard state of the 
same source for different outbursts are very similar 
(XTE J1550-564: Sobczak et al. 2000; Aquila X-1: Yu et al. 2003; 
4U 1705-44: Barret \& Olive 2002), the X-ray luminosity corresponding to 
the start of the hard-to-soft state transition is roughly described 
by either ASM rate or HEXTE rate at the peak of the `preceding' hard flare 
of an outburst (see Yu et al. 2003). 
Using the same spectral model for the same source 
(4U 1705-44 and Aql X-1: same model as used in Barret \& Olive 2002; 
XTE J1550-564: blackbody + cut-off power law + gaussian line at 6.4 keV), 
we have derived the unabsorbed peak X-ray flux at the peak of the 
hard flare. The flux, or the X-ray luminosity, 
can vary by a factor of 2.5. This suggests that the hard-to-soft 
state transition does not occur at a constant 
X-ray luminosity for the same source.  
  
\section{Discussion and Conclusion}
We have investigated the RXTE pointed observations of the 
outbursts of the black hole transient XTE J1550-564 and 
the neutron star transient Aquila X-1, together with those of 
the flares of the quasi-persistent neutron star LMXB 4U 1705-44. 
Two instances of a `preceding' hard X-ray flare were observed 
in each source (note: The outbursts discussed here are only those 
in which the source could reach the point of state transition in the outburst 
rise. The low amplitude low/hard state outbursts 
are out of the scope of this paper). We 
find that the HEXTE peak flux is proportional 
to the ASM peak flux, which the source reached a few days to 
several weeks later. We 
show that the X-ray luminosity corresponding to the hard-to-soft state 
transition can vary by a factor of more than two for the {\it same} source. 

If the X-ray luminosity is an indicator of the mass accretion rate 
on the outburst time scale, a varying X-ray luminosity of the 
hard-to-soft state transition shows that the transition 
is not solely determined by the mass accretion rate. 
We have demonstrated that for an individual source, the brighter the initial low/hard 
state of the outburst is, the brighter the later high/soft state will be. 
This is contrary to what is predicted by models that attribute 
the state transition to a change in a single mass accretion rate 
(Examples of such models are e.g Esin et al. 1998; 
Meyer, Liu \& Meyer-Hofmeister 2000). In the disk-instability model of 
soft X-ray transient outbursts, 
the energy reservoir of the X-ray outburst is the material accumulated in the 
accretion disk during `quiescence'. If an outburst consumes 
almost all the material in the disk, the outburst size 
should correlate with the disk mass. The more massive the disk is, the 
bigger the outburst will be, and the bigger both the hard and the soft 
flares will be. Disk mass provides the additional parameter 
that determines the state transition. If this also holds for 
the state transition during the outburst decay, it should occur at a lower 
X-ray luminosity than that during the rise because of a less 
massive disk (assuming the same physics applies). This agrees with the 
observations -- the so-called hysteresis effect 
(Miyamoto et al. 1995). 

%Alternative mechanisms might also 
%explain this. Mineshige (1996) suggests that 
%during such outburst decay, the evaporation of thin disk back to the ADAF 
%is slow and inefficient so that the soft-to-hard state 
%transition to occur at a lower X-ray luminosity than the hard-to-soft 
%state transition during the outburst rise. 

The correlation between the hard X-ray peak flux and the soft X-ray peak 
flux of the outbursts indicates that the outbursts are set up early. 
At the time when the hard-to-soft state transition occurs and the hard 
X-ray peak flux is known, the soft X-ray peak flux of the outburst 
can already be predicted as the 
entire outburst scales with the hard peak flux. 

The correlation between the hard X-ray 
peak flux and the soft X-ray peak flux requires a link between an 
earlier accretion flow/outflow which generates the hard 
X-ray flare and a later accretion flow which produces the soft X-ray 
outburst. In the single flow Comptonization models for the low/hard 
state which are composed of disk and corona (for a recent review 
see Nowak 2002), the correlation requires the hard X-ray photons 
to originate from inverse Comptonization of low energy photons 
in ultraviolet or optical from the disk flow further 
away from the central compact object, which produces the 
soft X-ray outburst peak at a later time when it propagates 
to the center. This requires future observations 
to confirm a correlation between the ultraviolet or optical flux 
and the hard X-ray flux in the rise of the low/hard state, as 
required by the inverse Comptonization models. On the other 
hand, our results supports a two-accretion-flow geometry 
at the beginning of a SXT outburst; 
at the start of the outburst rise, a non-disk flow arrives earlier 
from the outer disk and generates the hard X-ray flux, 
while the optically thick disk flow propagates inward on a viscous 
time scale (in a model involving a halo flow being an ADAF flow 
or a corona flow, the disk's propagation inward may be also related to 
a simultaneous collapse or contrast of the central ADAF or corona). The 
time lag between the two flows depends on the initial inner disk radius 
at the start of the outbursts. This is consistent with the model proposed 
by Chakrabarti \& Titarchuk (1995). A consideration of a sub-Keplerian corona 
flow originated from disk evaporation moving above the disk rather 
than comoving with the disk flow in the ADAF+disk+corona model (e.g. Meyer, 
Liu \& Meyer-hofmeister 2000) may also explain the observations.
 Smith, Heindl \& Swank (2002) discussed 
 supporting evidence for such two simultaneous, independent accretion 
flows in several black hole X-ray binaries. 
Our study is on the outburst rise, different to the count rate 
decrease discussed in Smith, Heindl \& Swank (2002), and includes both 
black hole and neutron star X-ray binaries. Furthermore, our 
study indicates that {\it 
the two accretion flows are related in \.{M}}; the 
two accretion flows are supplied with approximately {\it proportional} 
amounts of matter among the outbursts of the same source. It is worth 
noting that a similar two-related-accretion-flow geometry was used to 
interpret the `parallel tracks' in the plot of kHz QPO frequency and 
the X-ray flux in neutron star LMXBs (van der Klis 2002).   
As all the three sources in our study 
are known to be LMXBs, this relation may only apply 
to LMXB outbursts. 

The two-accretion-flow geometry suggests that the accretion state 
depends on which accretion flow dominates. 
If the observed accretion state 
is solely a consequence of the competition of the two flows, the critical X-ray 
luminosity corresponding to the appearance of the `propeller' 
regime for accreting neutron stars (Lamb, Pethick \& Pines 1973; 
Illarionov \& Sunyaev 1975) is not expected to be a 
constant but to vary with the composition of the two accretion flows. 
This implies that during the outburst rise, if the X-ray luminosity is 
below the level of the hard-to-soft state transition, the X-ray luminosity 
corresponding to the `propeller' state might relate to the `propeller' 
state of the halo flow; while during the outburst decay, if the X-ray 
luminosity is above the level of the soft-to-hard state transition, 
the X-ray luminosity corresponding to the `propeller' state might relate 
to the `propeller' state of the disk flow. The difference between 
the two accretion flows probably also introduces 
differences in the energy spectrum in the `propeller' state. 
The competition of the two accretion flows and the `propeller' 
mechanism may jointly contribute to the complexities of the spectral 
variation at low X-ray luminosity ranges 
(see e.g. color-intensity diagram of Aql X-1 shown 
in Muno, Remillard \& Chakrabarty 2002).
  
The association of the preceding hard X-ray flux with 
a halo flow rather than a disk flow can also put some 
constraints on the viscosity of the hot 
halo flow compared with the disk flow. The rise of 
the `preceding' hard flare from the `quiescent' level 
in this sources has a time scale of several days to several weeks (${T}_{1}$), 
while the rise of the soft X-ray outburst has a time scale 
of several weeks (${T}_{2}$). The ratio 
between ${T}_{2}$ and ${T}_{1}$ is about 2-3. 
Thus if the two-accretion-flow model is 
correct, then the non-disk flow is a sub-Keplerian flow 
rather than a free-fall flow. Assuming 
that ${T}_{2}$ and ${T}_{1}$ are related with the viscous time 
scales $t_{visc}\sim\frac{{R}^2}{\nu}$, and 
the viscosity of either the halo 
flow or the disk flow is similar among sources, ${T}_{2}$ and ${T}_{1}$ 
may reflect the size of the central hole in the disk 
and the disk size at the start of an outburst, respectively. 
   
Our results may also support the accretion geometry composed of an 
inner jet/outflow and an outer disk flow in which the jet/outflow generates 
the hard X-rays and the disk flow generates the soft X-rays 
(e.g. Markoff, Falcke \& Fender 2001). In this case, the correlation 
we found requires that the base of the jet is related to the disk flow 
and there is a close relationship between the peak powers of the two. 
The more massive the disk is, the stronger the jet will be (see also 
Garcia et al. 2003). 
This might suggest that the lifetime of the jet/outflow is more 
sensitive to the inner radius of the disk flow than 
the disk flow luminosity.  Fender \& Kuulkers (2001) show that 
soft X-ray peak luminosity also correlates with the radio peak luminosity. 
So our results predict that the radio peak flux correlates with the hard X-ray 
peak. 

\acknowledgments
The authors thank A. R. King for mentioning jet or outflow and 
Frederick K. Lamb for very useful comments. 
This work was supported mainly by the Netherlands Organization 
for Scientific Research (NWO) under grant 614.051.002 and partly by 
NASA grants NAG 5-12030 and NAG 5-8740, and NSF grant AST 0098399, 
and has made use of data obtained through the High 
Energy Astrophysics Science Archive Research Center Online Service, 
provided by the NASA/Goddard Space Flight Center.

\clearpage

\begin{figure*}
\epsscale{1.0}
\plottwo{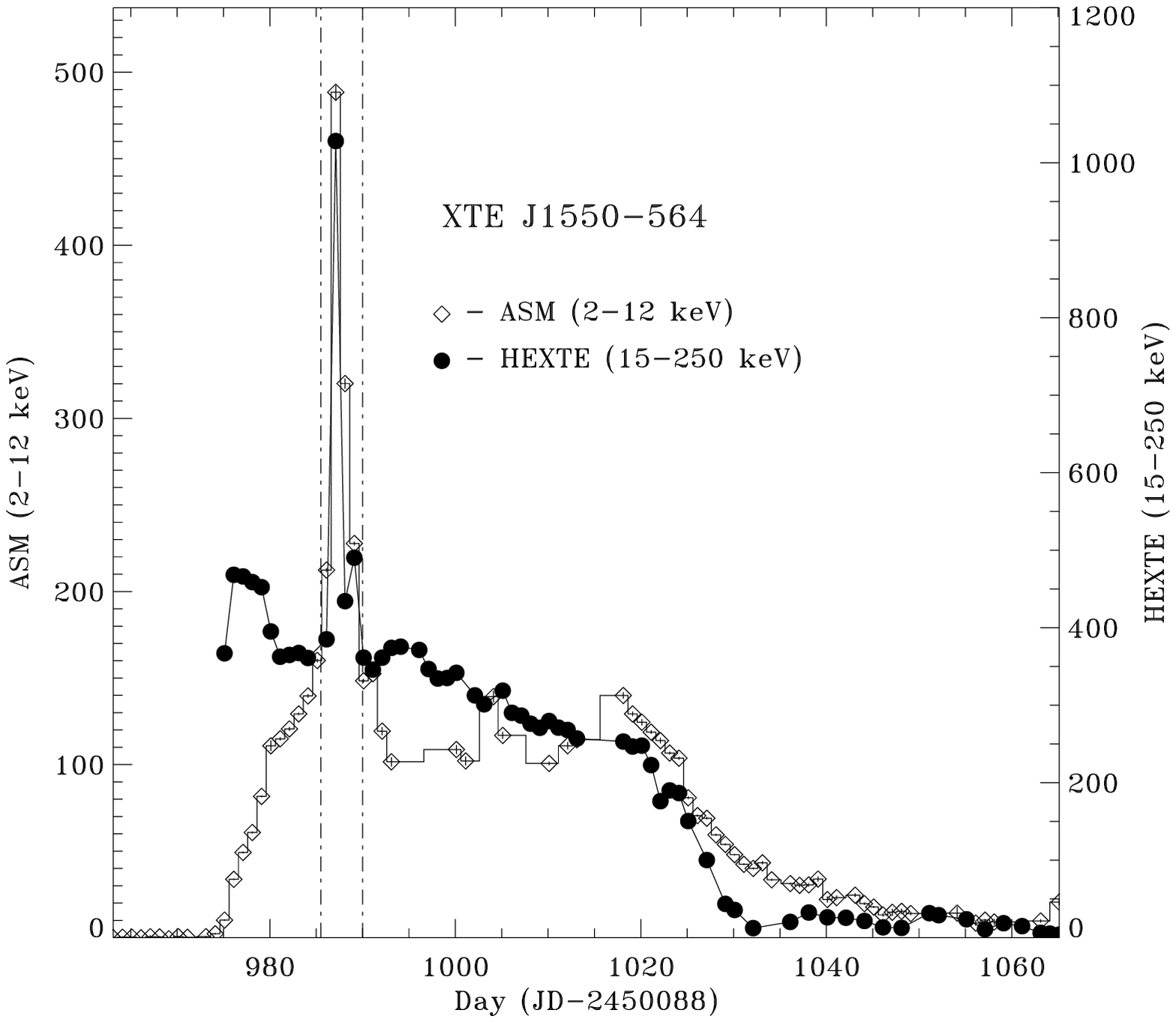}{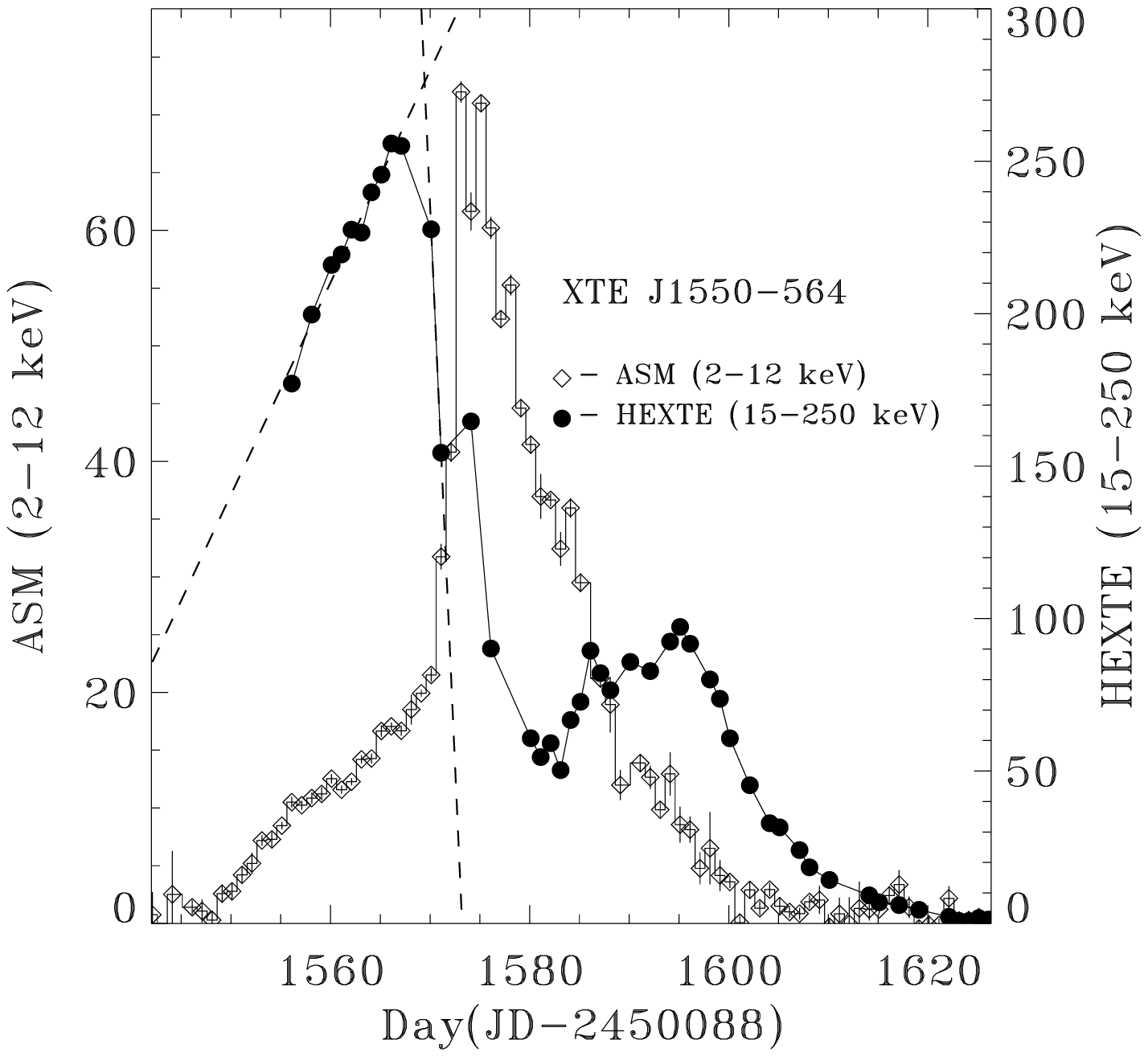}
\caption{The 1998-1999 outburst (left) and the 2000 May outburst (right) of XTE J1550-564. 
ASM light curves (2--12 keV) and the HEXTE light curves (15--250 keV) are shown
as diamonds and filled circles, respectively. The initial most bright low/hard
states are identified at the peaks of the `preceding' hard flares, corresponding
to Day 975 of 1998-1999 outburst and Day 1566 of 2000 outburst, respectively. It is 
worth noting that the RXTE observations covered full turnover of the `preceding' 
hard flare peak in the 1998-1999 outburst on a daily scale. For the 2000 outburst, 
we may miss RXTE pointings at the turnover of the hard peak. We 
extrapolate the slopes of the rise and the decay to derive an upper limit of the HEXTE peak 
flux, about 14 c/s higher than the peak rate observed ( $\sim$256 c/s).
  }
\end{figure*}

\begin{figure*}
\epsscale{1.0}
\plotone{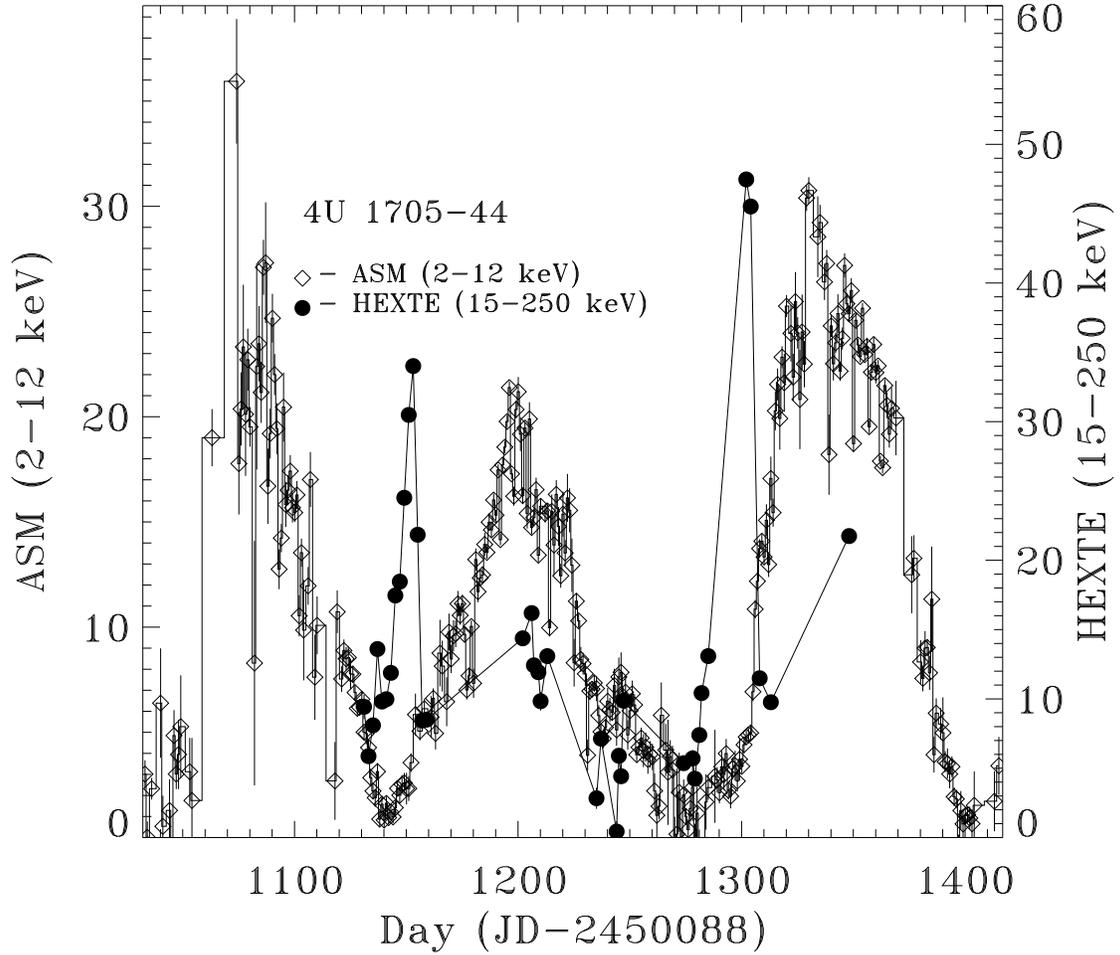}
\caption{The ASM light curves (2--12 keV) and the HEXTE light curves (15--250 keV) 
of 4U 1705-44 are shown
as diamonds and filled circles, respectively. It is 
worth noting that on a daily scale, the HEXTE pointings roughly covered the hard 
peak of the first flare but probably not the later. 
Notice that the ASM peaks lag the HEXTE peaks for 30--50 days, much longer than 
that observed in Aquila X-1 (Yu et al. 2003).}
\end{figure*}

\begin{figure*}
\epsscale{1.0}
\plotone{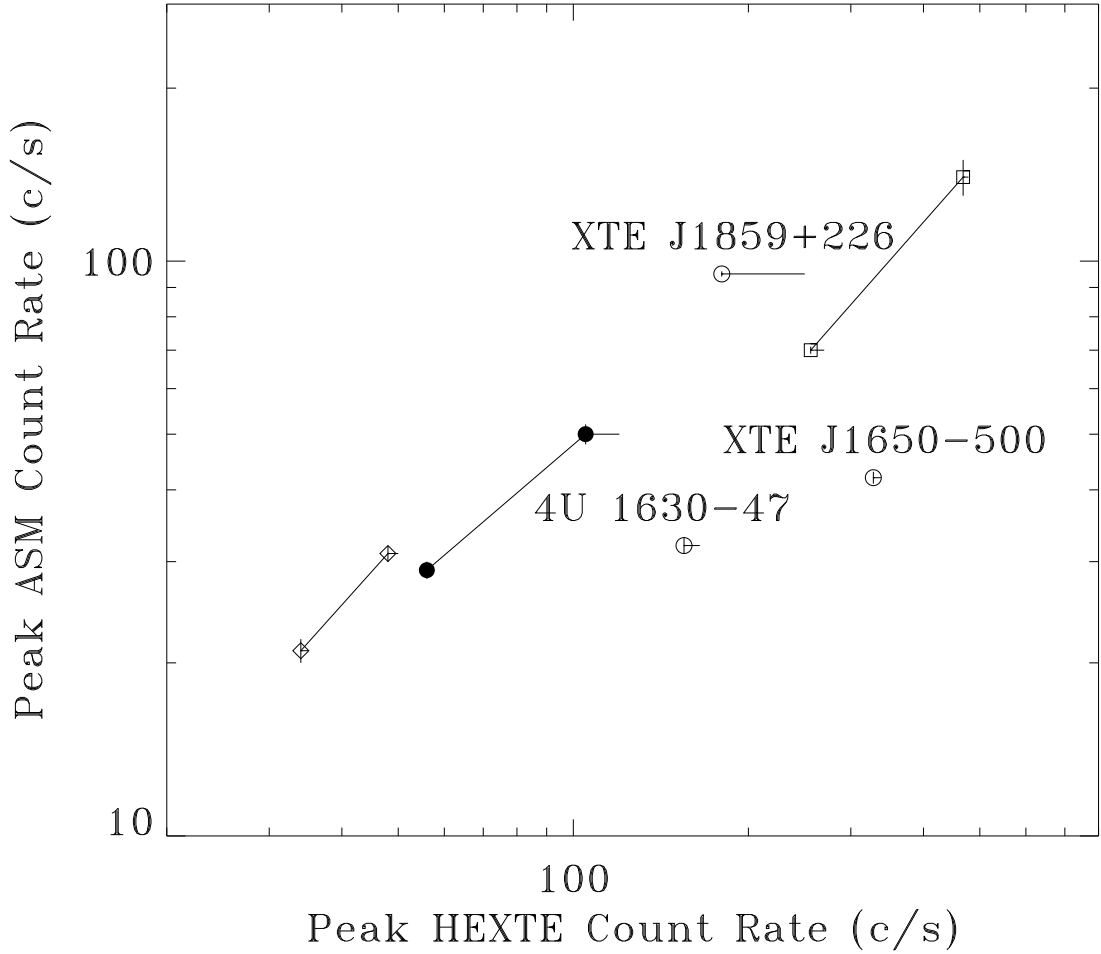}
\caption{The correlation between HEXTE (15-- 250 keV) 
and ASM (2--12 keV) peak flux in those outbursts or flares of 
XTE J1550-564 (squares), Aquila X-1 (filled circles), and 4U 1705-44 (diamonds) discussed 
in the text. The lack of coverage by the HEXTE pointed observations on a daily basis will introduce 
an underestimate of the peak HEXTE fluxes, but assuming a linear 
rise and decay, the effect of the underestimate is small. This is shown as error bars of 
the HEXTE peak flux. The outbursts in 4U 1630-47, XTE J1650-500, and XTE J1859+226 are also shown as open circles.
 }
\end{figure*}

\end{document}